\documentclass{amsart}

\usepackage{amsmath,amsfonts,amssymb}
\usepackage{algorithm,algorithmic}
\usepackage{hyperref}

\begin{document}

\title{Heisenberg Groups as Platform for the AAG key-exchange protocol}
\author{Delaram Kahrobaei, Ha T. Lam}

\maketitle

\begin{abstract}
Garber, Kahrobaei, and Lam studied polycyclic groups generated by number field as platform for the AAG key-exchange protocol. In this paper, we discuss the use of a different kind of polycyclic groups, Heisenberg groups, as a platform group for AAG by submitting Heisenberg groups to one of AAG's major attacks, the length-based attack.
\end{abstract}

After the introduction of the Anshel-Anshel-Goldfeld (AAG) key-exchange protocol  in 1999 \cite{AAG99}, it has been studied extensively with different groups as platform, for example, using braid groups by Ko et al. \cite{KoLee00}, using Thompson's group by Shpilrain and Ushakov \cite{ShpilrainUshakov05}. Different attack methods have also been applied to AAG \cite{lee2002potential, hofheinz2002practical, myasnikov2006random}.

One of the major attack methods of the AAG protocol is the length-based attack (LBA), originated with Hughes and Tannenbaum \cite{HughesTann02}, whose paper provided an example in braid groups, together with remark on the importance of the length function. Garber et al. studied the infeasibility of the length based attack with a choice of length function \cite{GarberLBA06}, but then introduced a variant of it using memory which succeeded in breaking AAG for braid group \cite{GarberProbabilistic05}. Myasnikov and Ushakov also studied the length-based attack for braid group and provided several variants with which it was possible to break AAG \cite{MyasnikovUshakov07}. Similar attack was implemented against system based on the Thompson group \cite{Thompson-LBA}

In 2004, Eick and Kahrobaei suggested using polycyclic groups as platform group for the AAG key-exchange protocol \cite{EickKahrobaei04}. This idea was realized by Garber, Kahrobaei, and Lam \cite{Lam13}. In that paper, several variants of LBA were tested on an AAG implementation using polycyclic groups generated from number field. The result suggests that this type of polycyclic group is resistant to the length-based attack. Taking inspiration from this, we want to study the Heisenberg group as platform group for AAG. We use the variants of the length-based attack presented in \cite{Lam13} to conduct tests on an implementation of the Heisenberg group. The result is then analysed, and we conclude that the Heisenberg groups can be used as platform for the AAG protocol given the correct parameters.

Furthermore, the conjugacy search problem has been used for several other cryptographic protocols, such as the non-commutative Diffie-Hellman key exchange \cite{KoLee00}, the non-commutative El-Gamal key exchange \cite{KK06}, the non-abelian Cramer-Shoup key exchange \cite{AK09} and the non-commutative digital signatures \cite{KK12}. The length-based attack can be applied to all of these protocols, hence testing different groups against it and collecting data about parameters that make them resistant to LBA is important, not just for the implementation of AAG but also of other protocols.

The paper is organized as follows: in Section~\ref{AAG}, we introduce the Anshel-Anshel-Goldfeld key exchange protocol, in Section~\ref{PolycyclicGroups}, there is a short review of polycyclic groups and the construction of Heisenberg group; in Section~\ref{LBA}, we talk about the length-based attack; and in Section~\ref{Results}, we detail the experiments, results and conclusion that we made.

\section{The Anshel-Anshel-Goldfeld Key Exchange Protocol and Heisenberg Group}
\label{AAG}
In this section, we give a short introduction to the Anshel-Anshel-Goldfeld key-exchange protocol. As usual, we use Alice and Bob as two parties who want to communicate over an insecure channel. The Anshel-Anshel-Goldfeld key exchange protocol works as follows:

Let $G=\langle g_1, \ldots, g_n \mid R \rangle$ be a finitely presented group with generators $g_1, \ldots, g_n$ and relation set $R$. First, Alice chooses, as her public set, $\overline{a}=(a_1,\ldots,a_{N_1})$ where $a_i \in G$ and Bob chooses, as his public set, $\overline{b}=(b_1,\ldots,b_{N_2})$ where $b_i \in G$. They both publish their sets. Alice then chooses her private key $A=a_{s_1}^{\varepsilon_1} \ldots a_{s_L}^{\varepsilon_L}$ where $a_{s_i} \in \overline{a}$ and $\varepsilon_i \in \{\pm1\}$. Bob also chooses his private key $B=b_{t_1}^{\delta_1} \ldots b_{t_L}^{\delta_L}$ where $b_{t_i} \in \overline{b}$ and $\delta_i \in \{\pm1\}$. Alice computes $b'_i=A^{-1}b_iA$ for all $b_i \in \overline{b}$ and sends it to Bob. Bob also computes $a'_i=B^{-1}a_iB$ for all $a_i \in \overline{a}$ and sends it to Alice. Now the shared secret key is $K=A^{-1}B^{-1}AB$. Alice can computes this key by
\begin{eqnarray}
K_A & = & A^{-1} a'^{\varepsilon_1}_{s_1} \cdots a'^{\varepsilon_L}_{s_L} = A^{-1} (B^{-1} a_{s_1} B)^{\varepsilon_1} \cdots (B^{-1} a_{s_L} B)^{\varepsilon_L} \nonumber \\
& = & A^{-1}B^{-1} a_{s_1}^{\varepsilon_1} \cdots a_{s_L}^{\varepsilon_L} B = A^{-1}B^{-1}AB = K \nonumber
\end{eqnarray}
Bob can likewise computes $K_B=B^{-1} b'^{\delta_1}_{t_1} \cdots b'^{\delta_L}_{t_L} = B^{-1}A^{-1}BA$, hence the shared key is $K=K_B^{-1}$.

In order to find $K$, the eaves-dropper needs to find either $A' \in \langle a_1, \ldots, a_{N_1} \rangle$ such that $\overline{b'}=A'^{-1}\overline{b}A'$ or find $B' \in \langle b_1, \ldots, b_{N_2} \rangle$ such that $\overline{a'}=B'^{-1}\overline{a}B'$. Thus, the security of AAG is based on the assumption that the subgroup-restricted simultaneous conjugacy search problem is hard.

\section{Polycyclic Groups}
\label{PolycyclicGroups}
In this section, we give a short review of polycyclic groups and discuss how we generate Heisenberg groups. 

\subsection{Polycyclic groups} Recall that $G$ is a \textit{polycyclic group} if it has a polycyclic series, i.e., a subnormal series $G=G_1 \rhd G_2 \rhd \ldots \rhd G_{n+1}=\{1\}$ with non-trivial cyclic factors. The polycyclic generating sequence of $G$ is the $n$-tuple $(g_1,\ldots, g_n)$ such that $G_i=\langle g_i,G_{i+1} \rangle$ for $1\leq i \leq n$.

Every polycyclic group has a finite presentation of the form:
$$\langle g_1,\ldots,g_n \mid g_j^{g_i}=w_{ij}, \; g_j^{g_i^{-1}}=v_{ij}, \; g_k^{r_k}=u_{kk} \; \text{for} \; 1\leq i<j\leq n \; \text{and} \; k \in I \rangle $$
where $w_{ij},v_{ij},u_{kk}$ are words in the generators $g_{i+1},\ldots ,g_n$ and $I$ is the set of $i \in \{1,\ldots,n\}$ such that $r_i=[G_i:G_{i+1}]$ is finite. Here $a^b$ stands for $b^{-1}ab$.

Using induction, we see that each element of $G$ defined by this presentation can be uniquely written as $g=g_1^{e_1} \ldots g_n^{e_n}$ with $e_i \in \mathbb{Z}$ for $1\leq i \leq n$, and $0 \leq e_i < r_i$ for $i \in I$. This is the \textit{normal form} of an element. If every element in the group can be presented uniquely in normal form, then the polycyclic presentation is \textit{consistent}. Note that every polycyclic group has a consistent polycyclic presentation. Thanks to the existence of normal form, the word problem in polycyclic groups can be solved efficiently, this is an important requirement of a platform group for AAG.

The \textit{Hirsch length} of a polycyclic group is the number of $i$ such that $r_i=[G_i:G_{i+1}]$ is infinite. This number is invariant of the chosen polycyclic sequence.

For more details regarding polycyclic groups, see \cite{Holt-handbook,Eick-habil}.

\subsection{Heisenberg groups}
Heisenberg groups have been studied widely from the point of view of analysis, geometry, physics, etc \cite{binzpods-book}. From the group theory point of view, they are often used as examples of nilpotent groups. For more discussion of Heisenberg groups and their group-theoretic properties, see \cite{epstein-book}.

The three dimensional Heisenberg group, often known as \textit{the} Heisenberg group, is the group of $3\times 3$ upper triangular matrices of the form $$\begin{pmatrix} 1&x&y\\0&1&z\\0&0&1 \end{pmatrix}$$ where $x,y,z \in \mathbb{R}$. Another presentation for it is $\langle a,b,c \mid [a,b]=c, [a,c]=[b,c]=1 \rangle$. The nilpotency for the Heisenberg group is easy to see, making it a polycyclic group.

Generalizing the Heisenberg group, we have higher dimension Heisenberg groups, $H^{2n+1}, n\geq 1, n \in \mathbb{Z}$. As matrix group, they are groups of dimension $n+2$ matrices of the form
$$\begin{pmatrix} 1&x_1&\ldots&x_n&c \\ 0&1&0&\ldots&y_n \\ \vdots&\vdots&\vdots&\vdots&\vdots \\
0&0&\ldots&1&y_1 \\ 0&0&\ldots&0&1 \end{pmatrix}$$
where $x_i,y_i,z \in \mathbb{R}$. The calculation for the commutator subgroup and the center of $H^{2n+1}$ is straight-forward, showing that it is also a nilpotent group, thus, also polycyclic. The Hirsch length of $H^{2n+1}$ is $2n+1$.

The Heisenberg groups of higher dimension $H^{2n+1}$ also has presentation $$\langle a_1,\ldots,a_n,b_1,\ldots,b_n,c \mid [a_i,b_i] = c, [a_i,c] = [b_i,c]=1, [a_i,a_j] =[b_i,b_j] = 1, i \neq j\rangle$$
This presentation makes it easy to encode $H^{2n+1}$ in a computer system. In fact, the GAP (Groups, Algorithms, Programming) system \cite{GAP4} can compute $H^{2n+1}$ as part of the polycyclic package by Eick and Nickel \cite{GAPPolycyclic}. Using this implementation of $H^{2n+1}$, we study the Heisenberg groups as platform group for the AAG key-exchange protocol, in particular, under the length-based attack, one of the major attacks for AAG.

\section{Length Based Attack}
\label{LBA}
\subsection{Overview of the length-based attack}
The length-based attack is a probabilistic attack against AAG, with the goal of finding Alice's (or Bob's) private key. It is based on the idea that a conjugation of the right element will decrease the length of the captured package. Using notation of Section ~\ref{AAG}, the captured package is $\overline{b'}=(b'_1, \ldots , b'_{N_2})$ where $b'_i=A^{-1}b_iA$. If we conjugate $\overline{b'}$ with elements from the group $\langle a_1, \ldots, a_{N_1} \rangle$ and the resulting tuple has decreased length, then we know we have found a conjugating factor. The process of conjugation is then repeated with the decreased length tuple until another conjugating factor is found. The process ends when the conjugated captured package is the same as $\overline{b}=(b_1, \ldots , b_{N_2})$, which is public knowledge. Then the conjugate can be recovered by reversing the sequence of conjugating factors. The idea of the length-based attack can be summarized as going from bottom to top of the tower:

\begin{center}
$b_i$ \\ $\downarrow$ \\ $a_{s_1}^{-\varepsilon_1} b_i a_{s_1}^{\varepsilon_1}$  \\ $\downarrow$ \\ $a_{s_2}^{-\varepsilon_2} a_{s_1}^{-\varepsilon_1} b_i a_{s_1}^{\varepsilon_1} a_{s_2}^{\varepsilon_2}$ \\ $\downarrow$ \\ $\vdots$ \\ $\downarrow$ \\ $a_{s_L}^{-\varepsilon_L} \ldots a_{s_2}^{-\varepsilon_2} a_{s_1}^{-\varepsilon_1} b_i a_{s_1}^{\varepsilon_1} a_{s_2}^{\varepsilon_2} \ldots a_{s_L}^{\varepsilon_L}$
\end{center}

For more details on the length-based attack, see \cite{Shpilrain-book,MyasnikovUshakov07}.

\subsection{LBA with memory 2}
In \cite{GarberLBA06,GarberProbabilistic05,MyasnikovUshakov07,Thompson-LBA, Lam13}, several variations of the length-based attack are given. Here we recall Algorithm 4 of \cite{Lam13} which we use in our experiments.

In this algorithm, $S$ holds $M$ tuples every round and all elements of $S$ are conjugated, but only the $M$ smallest conjugated tuples (by total length) are added back into $S$. Because we are adding back $M$ smallest tuples and not just one single element, we avoid the problem of the same element being removed and added time and again. Moreover, this method of saving several tuples, not just the ones whose length decreased after conjugation, keeps us away from problems generated by peaks. Here, for the stopping condition, we use a time-out that is defined by the user.

\begin{algorithm}
\caption{LBA with Memory 2}
\label{algoMem2}
\begin{algorithmic}[1]
	\STATE Initialize $S=\{(|\overline{b'}|,\overline{b'},{id}_G)\}$.
	\WHILE{not time out}
		\FOR{$(|\overline{c}|,\overline{c},x) \in S$}
			\STATE Remove $(|\overline{c}|,\overline{c},x)$ from $S$
			\STATE Compute $\overline{c}^{a_i^\varepsilon}$ for all $i \in \{1 \ldots N_1\}$ and $\varepsilon=\pm 1$
			\STATE \textbf{if} $\overline{c}^{a_i^\varepsilon}=\overline{b}$ \textbf{then} output inverse of $xa_i^\varepsilon$ and stop
			\STATE Save $(|\overline{c}^{a_i^\varepsilon}|,\overline{c}^{a_i^\varepsilon},xa_i^\varepsilon)$ in $S'$
		\ENDFOR
		\STATE After finished all conjugations, sort $S'$ by the first element of every tuple
		\STATE Copy the smallest $M$ elements into $S$ and delete the rest of $S'$
	\ENDWHILE
	\STATE Otherwise, output FAIL
\end{algorithmic}
\end{algorithm}

\section{Results}
\label{Results}
As a test of the resilience of polycyclic groups against the length-based attack, Garber, Kahrobaei, and Lam \cite{Lam13} implemented four variants of the length-based attack and performed experiments against all four variants. The conclusion was that Algorithm 4, LBA with Memory 2, was the most effective. Hence, in this study, we use Algorithm 4 for all of our experiments.

We performed several sets of tests, all of which were run on an Intel Core I7 quad-core 2.0GHz computer with 12GB of RAM, running Ubuntu version 12.04 with GAP version 4.5 and 10GB of memory allowance. In all these tests, the Heisenberg group $G$ with $2n$ generators having Hirsch length $h(G)=2n+1$ is generated using GAP with the Polycyclic group package \cite{GAPPolycyclic}. The size of Alice’s and Bob’s public sets are both $N_1=N_2=20$, the memory used is $M=1000$ and the time-out is 30 minutes.

To see the effect of element length, we fix the key length $L=10$, but changes the range of random element between $[L_1,L_2]=[10,13]$ and [20,23]. The result is as follows:
\begin{center}
\begin{tabular}{|c|c|c|c|}
\hline
 n & h(G) & $[L_1,L_2]=[10,13]$ & $[L_1,L_2]=[20,23]$\\
\hline
 5 & 11 & 29\% & 53\% \\
 6 & 13 & 69\% & 39\% \\
 7 & 15 & 51\% & 58\% \\
 8 & 17 & 62\% & 67\% \\
\hline
\end{tabular}
\end{center}
As we can see, changing the length of random element does not have a dramatic effect on the success rate. Hence, to ensure a lower rate of success, one should look for other factors like the Hirsch length or the key length.

To see the effect of Hirsch length, we use a small key length $L=10$ to increase the possibility of success. The length of each random element is in $[L_1,L_2]=[10,13]$. The result is as follows:
\begin{center}
\begin{tabular}{|c|c|c|c|}
\hline
 n & h(G) & Time & Success rate \\
\hline
 3 & 7 & 45.42 hours & 11\% \\
 5 & 11 & 37.82 hours & 29\% \\
 6 & 13 & 21.12 hours & 69\% \\
 7 & 15 & 28.95 hours & 51\% \\
 8 & 17 & 24.33 hours & 62\% \\
\hline
\end{tabular}
\end{center}
Curiously, the higher the Hirsch length in this case, the higher and quicker the length-based attack succeeded. This interesting result warrant further investigations.

To illustrate the effect of key length, in the following experiment, we let the length of each random element to be in $[L_1,L_2]=[20,23]$ and vary the key length $L$.
\begin{center}
\begin{tabular}{|c|c|c|c|c|}
\hline
 n & h(G) & L=10 & L=20 & L=50\\
\hline
 5 & 11 & 53\% & 11\% & 1\% \\
 6 & 13 & 39\% & 7\% & 0\% \\
 7 & 15 & 58\% & 5\% & 1\% \\
 8 & 17 & 67\% & 9\% & 7\% \\
\hline
\end{tabular}
\end{center}
Clearly, increasing the key length reduced the success rate dramatically. What is interesting is that the same algorithm, together with the same parameters, when applied on polycyclic group of Hirsch length 10 generated with number field as in \cite{Lam13} gave 0\% success rate. However, it is clear that we need different parameters to ensure low success rates.

We turn the key length to $L=50$, let length of random element be $[L_1,L_2]=[40,43]$. Note that this is the parameters used in \cite{MyasnikovUshakov07}, which succeeded in breaking AAG with braid group.
\begin{center}
\begin{tabular}{|c|c|c|c|}
\hline
 n & h(G) & L=50\\
\hline
 6 & 13 & 0\% \\
 7 & 15 & 0\% \\
 8 & 17 & 1\% \\
\hline
\end{tabular}
\end{center}
As expected, this produces almost zero success rate. We recommend this as parameters for Heisenberg group as platform for AAG.

Finally, as a stress test, we increase the time-out to 4 hours instead of the usual 30 minutes per test. Because of the long time-out, we only do 50 tests each, instead of 100 test batch as in the previous experiments. The element length is kept at $[L_1,L_2]=[20,23]$, key length is $L=20$.
\begin{center}
\begin{tabular}{|c|c|c|c|}
\hline
 n & h(G) & L=20\\
\hline
 6 & 13 & 3\% \\
 7 & 15 & 4\% \\
 8 & 17 & 7\% \\
\hline
\end{tabular}
\end{center}
Even at such long time-out, success rate is quite modest, given that the key length we use here is quite small. 

With these results, we conclude that the Heisenberg groups work well as platform for the AAG protocol given the correct parameters. This strengthens the idea to use polycyclic group as platform for AAG in particular, and non-commutative cryptographic primitives in general.

\bibliographystyle{plain}
\bibliography{AAGPlatformBibtex.bib}
\end{document}